\documentclass[a4paper,english,12pt]{article}

\usepackage{hyperref}

\usepackage{fontenc}
\usepackage[latin1]{inputenc}
\usepackage{float}
\usepackage{color}
\usepackage{graphicx}
\usepackage{amssymb}
\usepackage{babel}
\usepackage{appendix}
\usepackage{amsmath}
\usepackage{fancyhdr}
\usepackage[affil-it]{authblk}
\usepackage{geometry}
\begin{document}

\title{On the Excitation and Radiative Decay Rates of Plasmonic Nanoantennas}

\author[1]{Kalun Bedingfield}
\author[1]{Angela Demetriadou\thanks{a.demetriadou@bham.ac.uk}}

\affil[1]{School of Physics and Astronomy, University of Birmingham, Edgbaston, Birmingham B15 2TT, United Kingdom}

\date{}
\maketitle

\abstract{Plasmonic nanoantennas have the ability to confine and enhance incident electromagnetic fields into very sub-wavelength volumes, while at the same time efficiently radiating energy to the far-field. These properties have allowed plasmonic nanoantennas to be extensively used for exciting quantum emitters---such as molecules and quantum dots---and also for the extraction of photons from them for measurements in the far-field. Due to electromagnetic reciprocity, it is expected that plasmonic nanoantennas radiate energy as efficiently as an external source can couple energy to them. In this paper, we adopt a multipole expansion (Mie theory) and numerical simulations to show that although reciprocity holds, certain plasmonic antennas radiate energy much more efficiently than one can couple energy into them. This work paves the way towards designing plasmonic antennas with specific properties for applications where the near-to-far-field relationship is of high significance, such as: surface-enhanced Raman spectroscopy, strong coupling at room temperature, and the engineering of quantum states in nanoplasmonic devices.}


\section{\label{sec:Intro}Introduction}
Isolated metallic nanoparticles (NPs) produce large local field enhancements via the excitation of localised plasmons, and can efficiently radiate energy to the far-field. Plasmonic nanoantennas are usually composed of two or more tightly coupled metallic nanostructures and can concentrate electromagnetic fields to even smaller nanoscale `hot-spots', enhancing the light intensity by at least three orders of magnitude~\cite{Akselrod2014,Sugimoto2018,Baumberg2019,Hoang2015,Yashima2016,Huang2019, Kishida2022}. 
During the last few years, there have been tremendous advancements in the fabrication of plasmonic nanoantennas, with gaps reaching just few (or even sub) nanometers---often referred to as plasmonic nanocavities~\cite{Sigle2015,Benz2015,Emboras2016}. 
Plasmonic nanocavities produce extremely sub-wavelength confinement of light, which has led to unique and extraordinary realizations, such as: room temperature strong coupling of a single molecule~\cite{Chikkaraddy2016a,Zengin2015,Hoang2016}, imaging of a single molecule~\cite{Zhang2017,Yang2020}, and even the formation of `pico-cavities' to interrogate specific chemical bonds within a single molecule~\cite{FelixBenz2016}.

\begin{figure} 
\includegraphics[width=\linewidth]{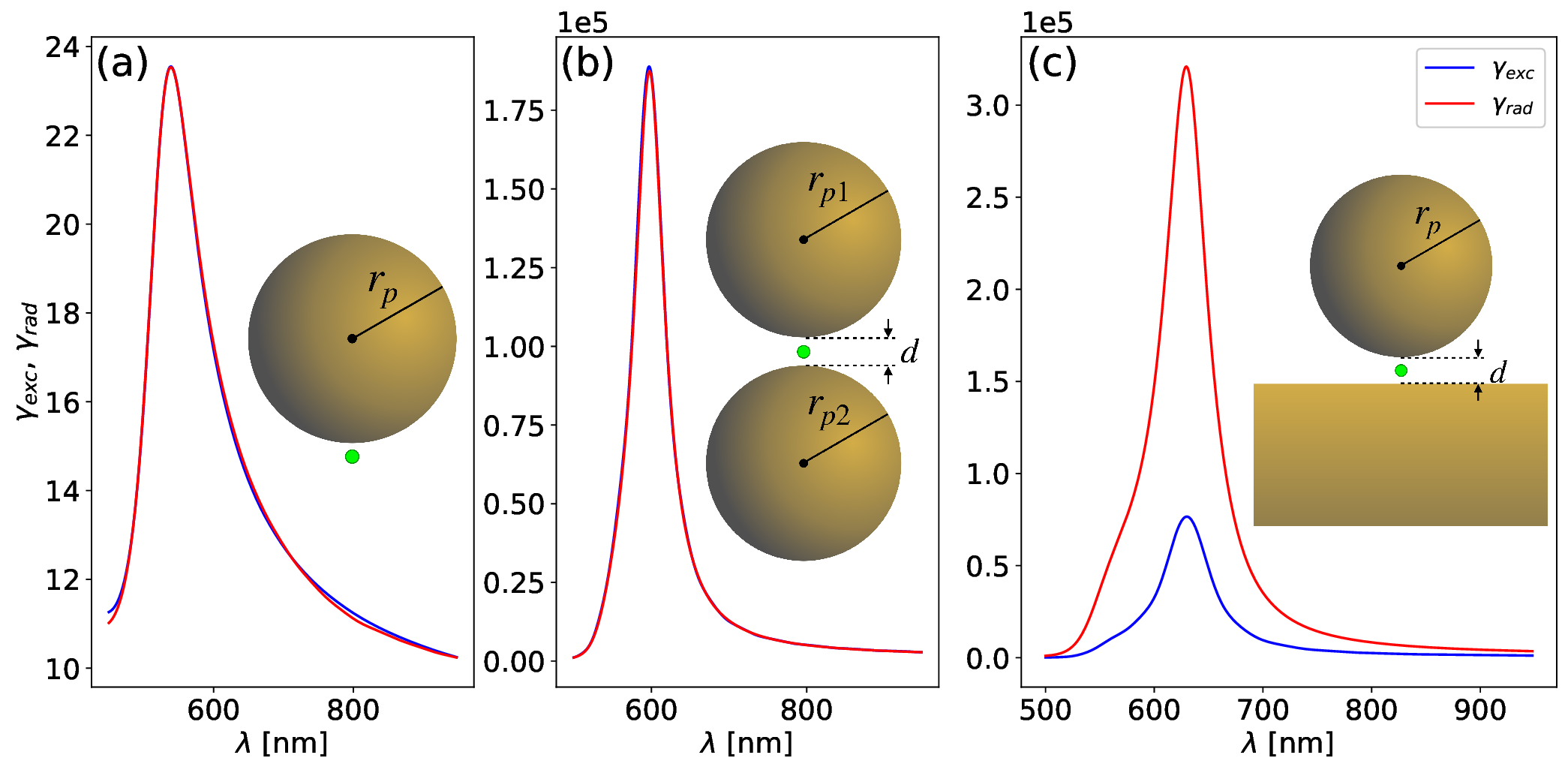}
\caption{\label{fig:NanoCavYComp} The $\gamma_{exc}$ (blue) and $\gamma_{rad}$ (red) obtained from FDTD calculations for different nanoplasmonic systems: (a) Isolated $2r_p = 60$nm gold NP. (b) Dimer antenna of $2r_{p1} = 2r_{p2} = 60$nm gold NPs with $d=1$nm separation and (c) NPoM geometry of a $2r_p = 60$nm gold NP assembled $d=1$nm above a gold substrate. The green dot indicates the position of the dipole source when determining the $\gamma_{rad}$, and the location where the fields are measured when calculating the $\gamma_{exc}$.}
\end{figure}

All of the aforementioned recent advances utilize the strong near-field enhancement to produce excitations that are large enough to be emitted via the plasmon and measured experimentally in the far-field. 
Most experimental and theoretical studies of plasmonic nanocavities have focused on scattering methods that represent the resonant modes of the cavity in the far-field, with very few studies focusing on the near-field enhancement, modal decomposition~\cite{Kongsuwan2020} and radiative efficiencies---sometimes with unexpected results~\cite{Li2020}. 
Far-field spectra offer limited information on how quantum emitters (QEs)---such as fluorescent molecules, quantum dots and analytes---in a nanocavity experience and interact with the near-field enhancement; and much less information on how the energy, photons and molecular Raman signals radiate energy out of the nanocavity via the plasmons to be detected in the far-field. 
It is often thought that energy couples into plasmonic devices from the far-field (in-coupling) as efficiently as a QE in the nanocavity radiates to the far-field (out-coupling), since both processes occur via the same set of plasmonic modes. 
In this paper, we show that the above statement does not always hold, even though electromagnetic reciprocity conditions always remain satisfied. We explain the origin of this behaviour by adopting a multipolar decomposition---using the Green's tensor description of Mie theory---to determine what type of plasmonic nanoantennas demonstrate this unexpected behaviour. 
For simplicity, we focus on three representative plasmonic systems: (i) isolated spherical NPs; (ii) various dimer antennas, consisting of two tightly coupled spherical NPs; and (iii) the NanoParticle on Mirror (NPoM) configuration, formed of a spherical NP assembled on a flat metal surface---all three geometries are shown in the insets of Figure~\ref{fig:NanoCavYComp}. 

The properties of plasmonic nanoantennas as an environment for a QE are often characterized by measuring the rate with which an external source would excite the QE in the antenna (i.e. excitation rate, $\gamma_{exc}$), and the rate at which a dipole source within the antenna radiates energy to the far-field (i.e. radiative decay rate, $\gamma_{rad}$).
If a molecule/QE is placed within the nanoantenna at position $\mathbf{r'}$, the plasmonic environment enhances its $\gamma_{exc}$ as:
\begin{equation}
\gamma_{exc} = \frac{|\hat{\mathbf{\mu}} \cdot \mathbf{E}(\mathbf{r}')|^2}{|\hat{\mathbf{\mu}} \cdot \mathbf{E}_0(\mathbf{r}')|^2}  \label{eq:excitation_rate}
\end{equation}
where $\hat{\mathbf{\mu}}$ is a unit vector parallel to the dipole moment of the molecule at position $\mathbf{r'}$; and $\mathbf{E}(\mathbf{r'})$, $\mathbf{E}_0(\mathbf{r'})$ are the total electric fields at position $\mathbf{r'}$ with and without the plasmonic system present respectively~\cite{LucasNovotny,Anger2006a}.  
On the other hand, the radiative behaviour of the plasmonic system---described by $\gamma_{rad}$---measures how efficiently a QE placed in the plasmonic antenna at position $\mathbf{r'}$ emits energy to the far-field. 
By definition this is calculated by considering a classical dipole source at position $\mathbf{r'}$ within the plasmonic antenna, integrating the total energy emitted by the combined system that crosses a surface enclosing the system, and normalizing with the energy of the dipole source~\cite{Ruppin1982a}. 

Due to reciprocity, electromagnetic systems emit and receive energy via the same modes; a signal must therefore remain unchanged if a source and a detector are inter-changed~\cite{Carminati1998,Porto2000,LucasNovotny}. 
Mathematically one can express this using Green's functions, with reciprocity dictating: $ G(\mathbf{r},\mathbf{r'})=G(\mathbf{r'},\mathbf{r})$, where the first and second arguments of the Green's function refer to the locations of the detector and source, respectively. 
This is an universal property of electromagnetics, including plasmonics. 
One would therefore expect the rate at which energy can be coupled into and out of the system to be equal. 

In Figure~\ref{fig:NanoCavYComp}(a), we show the $\gamma_{exc}$ and $\gamma_{rad}$ for an isolated gold NP of diameter $2r_p = 60$nm, numerically calculated using Finite-Difference Time-Domain (FDTD) methods~\cite{FDTD}, which indeed shows that $\gamma_{exc}=\gamma_{rad}$. 
Now we consider two such NPs of diameters $2r_{p,1}=2r_{p,2}=60$nm and place them close together to form a gap of just $d=1$nm, therefore creating a symmetric dimer antenna. 
The plasmonic modes of each NP couple with the modes of the other, they hybridise to produce strong field enhancements in the gap, and red-shift the overall plasmonic resonances. 
Despite the significant changes to the system and its plasmon modes, the $\gamma_{exc}$ and $\gamma_{rad}$ remain equal---as shown in Figure~\ref{fig:NanoCavYComp}(b).
A very similar plasmonic system is that of the NPoM antenna; this consists of a gold NP assembled above a flat gold substrate, forming a nanoscale gap. 
Here, the NP has diameter $2r_p = 60$nm and the gap is $d=1$nm to allow comparisons with the dimer antenna. 
The $\gamma_{exc}$ and $\gamma_{rad}$ for the NPoM antenna are shown in Figure~\ref{fig:NanoCavYComp}(c), where one can see that they exhibit a very different behaviour, with $\gamma_{exc}\neq\gamma_{rad}$.
For the NPoM antenna, the $\gamma_{rad}$ is one order of magnitude larger than the $\gamma_{exc}$, which means that a QE in the cavity is able to radiate energy to the far-field much more efficiently than an external wave can excite it. 
This result is surprising not only because it appears to violate reciprocity, but as the NPoM and dimer antennas are very often considered to be near identical plasmonic systems---due to the image charges that form in the mirror of the NPoM. 

The fact that the radiative rate of the NPoM is much stronger than its excitation rate has serious consequences: contributing to the suppressed quenching~\cite{Kongsuwan2018a},  the realization of single molecule strong coupling at room temperature~\cite{Chikkaraddy2016a}, and the efficient mapping of cavity hot-spots via the Raman response of molecules~\cite{Chikkaraddy2018} to name but a few. 
Previous work~\cite{Kongsuwan2018a} has shown that the mode confinement in the gap changes the dark nature of higher order modes to bright, which leads to stronger radiative emissions ($\gamma_{rad}$) for both the dimer antenna and the NPoM.  
This is also evident from our results in Figure~\ref{fig:NanoCavYComp}, where both the $\gamma_{exc}$ and $\gamma_{rad}$ increase by 4 orders of magnitude for both the dimer antenna and NPoM compared to the isolated NP.
This also leads to an increased quantum yield by 3 orders of magnitude as shown in Figure~S1(d-f).
However, the NPoM system shows even stronger $\gamma_{rad}$, and most importantly that $\gamma_{rad}\neq\gamma_{exc}$.
In this paper, we identify the origin of this behaviour and show how to design plasmonic nanoantennas with unique and tailored in- and out-coupling properties.

\section{\label{sec:Decomp}Modal Decomposition for the Excitation and Radiative Decay Rates}
To better understand the origin of this unusual behaviour for the NPoM cavity, Mie theory is used to perform a multipolar decomposition for the modes supported by isolated NPs. 
It can obtain mathematical expressions and therefore a more intuitive understanding for the plasmons on any sized NP. 
We consider an isolated spherical NP that is excited by a dipole source, placed at position $\mathbf{r'}$~\cite{Ruppin1982a,Bereza2017,Ohtaka1982,Chewb,GarciADeAbajoa,GarciADeAbajo,Yan,Wvatr1962,Kerker1980,Zhang2019,Low1997,Chew1987}. The Green's dyadic tensor for the dipole source fields in free space is given by~\cite{Tai1994}: 
\begin{eqnarray}
\mathbf{\overleftrightarrow{G}}_{inc} (\mathbf{r},\mathbf{r'}) =  \sum_{l,m}\sum_{e,o} C_{l,m}
\begin{cases}
\mathbf{M}^{e,o \ (1)}_{l,m} (k\mathbf{r'})\otimes  \mathbf{M}^{e,o \ (3)}_{l,m} (k\mathbf{r}) + \mathbf{N}^{e,o \ (1)}_{l,m} (k\mathbf{r'})\otimes \mathbf{N}^{e,o \ (3)}_{l,m} (k\mathbf{r}) \ \ \ \ r > r' \\
\mathbf{M}^{e,o \ (3)}_{l,m} (k\mathbf{r'}) \otimes \mathbf{M}^{e,o \ (1)}_{l,m} (k\mathbf{r}) + \mathbf{N}^{e,o \ (3)}_{l,m} (k\mathbf{r'}) \otimes \mathbf{N}^{e,o \ (1)}_{l,m} (k\mathbf{r}) \ \ \ \ r < r'
\end{cases} \label{eq:Greens_dyadic}
\end{eqnarray}
where $C_{l,m} = \frac{ik}{4\pi} (2 - \delta_0) \frac{2l + 1}{l(l+1)} \frac{(l-m)!}{(l+m)!}$ with $\delta_0=
\left\{
\begin{array}{lr} 
1, \ \ \ & m=0 \\ 
0, \ \ \ & m\neq 0
\end{array}
\right.
$, $k$ is the wavevector, and $\mathbf{M}^{e,o}_{l,m}$, $\mathbf{N}^{e,o}_{l,m}$ are the vector spherical harmonics $\mathbf{M}^{e,o}_{l,m}= \nabla \times \left( \mathbf{r} \psi^{e,o}_{l,m} \right) $ and $\mathbf{N}^{e,o}_{l,m} = \frac {1}{k} \nabla \times \mathbf{M}^{e,o}_{l,m} $---given in full form in Supp. Info. Section S2---obtained from the scalar wavefunctions: $\psi^{e,o}_{l,m} (\mathbf{r})= z_l(kr) P_l^m\left(\cos\theta\right) \left\{ 
\begin{array}{lr} 
&\cos m\phi \\ 
&\sin m\phi 
\end{array}
\right. $, which are the even and odd solutions to the homogeneous scalar Helmholtz equation~\cite{Tai1994,Bohren1983}. 
The superscripts $^{(1)}$ and $^{(3)}$ respectively refer to the use of spherical Bessel functions of the first ($j_{l}(kr)$) and third (Hankel, $h_{l}^{(1)}(kr)$) kinds for the general Bessel function $z_l(kr)$ in $\psi^{e,o}_{l,m}$. 
From the Green's dyadic, the electric fields can be obtained via $\mathbf{E}(\mathbf{r}) = \omega^2 \mu_0  \mathbf{\overleftrightarrow{G}} (\mathbf{r},\mathbf{r'}) \cdot \mathbf{p}(\mathbf{r'})$, where $\mathbf{p}(\mathbf{r'}) = \mathbf{p_0} \delta(\mathbf{r}-\mathbf{r'})$ is the dipole moment $\mathbf{p_0}$ of the emitter placed at $\mathbf{r'}$~\cite{LucasNovotny}. 
Note that similar formalisms to Eq.~\ref{eq:Greens_dyadic} hold for the fields transmitted inside the spherical NP, as well as for those scattered from its surface (see Supp. Info. Section~2 for the full mathematical description). 
After applying the boundary conditions at the surface of the NP and assuming we have non-magnetic materials (i.e. $\mu_{1}=\mu_{2}=1$), the Mie scattering and internal coefficients emerge~\cite{Tai1994,Bohren1983}. 
Here, we state only the two scattering coefficients:
\begin{eqnarray}
a_l & = & \frac{N^2 j_l(Nkr_p) [kr_p j_l(kr_p)]' - j_l(kr_p) [Nkr_p j_l(Nkr_p)]'}{N^2 j_l(Nkr_p) [kr_p h^{(1)}_l(kr_p)]' - h^{(1)}_l(kr_p) [Nkr_p j_l(Nkr_p)]'} \nonumber \\
b_l  & = & \frac{j_l(Nkr_p) [kr_p j_l(kr_p)]' - j_l(kr_p) [Nkr_p j_l(Nkr_p)]'}{j_l(Nkr_p) [kr_p h^{(1)}_l(kr_p)]' - h^{(1)}_l(kr_p) [Nkr_p j_l(Nkr_p)]'}
\end{eqnarray}
where the NP has refractive index $N$ and radius $r_p$, and is placed in a vacuum. 
Hence, the fields scattered from the spherical NP due to a dipole source at $\mathbf{r}'$ are given by: 
\begin{equation}
\mathbf{E}_{scat}(\mathbf{r})= \omega^2 \mu_0 \sum_{L} C_{l,m} \left[ b_l \ s_{l,m}^{e,o \ (3)} (k\mathbf{r'})  \mathbf{M}^{e,o \ (3)}_{l,m} (k\mathbf{r})  + a_l \ t_{l,m}^{e,o \ (3)}(k\mathbf{r'})  \mathbf{N}^{e,o \ (3)}_{l,m} (k\mathbf{r}) \right]  \label{eq:scatter-field}
\end{equation}
where $\sum_L=\sum_{l,m} \sum_{e,o}$, and we introduce the abbreviated notation~\cite{Ruppin1982a}: $s_{l,m}^{e,o \ (i)} (k\mathbf{r'})  = \mathbf{M}^{e,o \ (i)}_{l,m} (k\mathbf{r'}) \cdot  \mathbf{p}(\mathbf{r'})$ and $t_{l,m}^{e,o \ (i)}(k\mathbf{r'})  = \mathbf{N}^{e,o \ (i)}_{l,m} (k\mathbf{r'}) \cdot  \mathbf{p}(\mathbf{r'})$, which are scalar coefficients that describe how energy couples into modes due to a source at $\mathbf{r'}$ with dipole moment $\mathbf{p}(\mathbf{r'})$. 
The second term in the parenthesis of Eq.~\ref{eq:scatter-field} describes the electric (TM) modes (for metallic NPs, these correspond to the plasmonic modes), and the first term describes the magnetic (TE) modes (for metallic NPs, these are negligibly small). 

\begin{figure}
\includegraphics[width=\linewidth]{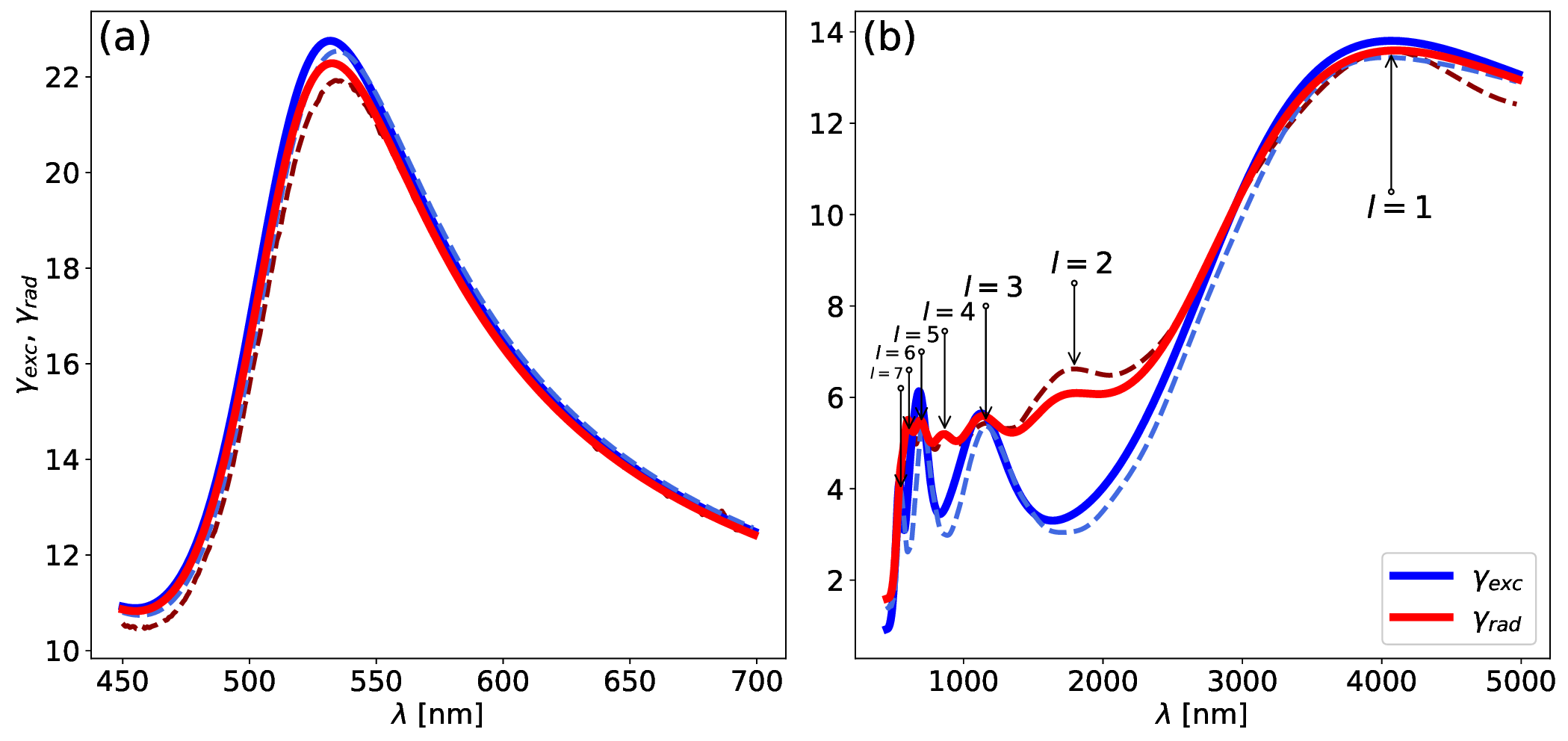}
\caption{\label{fig:IsoYYComp} The $\gamma_{exc}$ (blue) and $\gamma_{rad}$ (red) for isolated spherical NPs of diameters (a) $60$nm and (b) 1$\mu$m.
The analytical multipolar decompositions (full lines) and numerical calculations (dashed lines) are plotted together for comparison. In (b), the series of resonances are labelled by their corresponding angular momentum quantum number, $l$, which emerges from the multipolar decomposition.}
\end{figure}

The $\gamma_{exc}$ measures the energy coupled into the plasmonic system from a far-field source (i.e. plane wave). Using the above formalism for an $x$-polarized dipole moment---along with Eq.~\ref{eq:excitation_rate}---and taking the limit where $\mathbf{r'}\rightarrow \left[r'=\infty,\theta'=\pi,\phi'=0\right]$ (which corresponds to a dipole source placed at $z'=-\infty$) creates an $x$-polarized plane wave propagating along the positive $z$-axis. 
This reduces the incident and scattered fields of Eq.~\ref{eq:Greens_dyadic} and~\ref{eq:scatter-field} to~\cite{Bohren1983}: 
\begin{equation}
\begin{pmatrix}
\mathbf{E}_{PW}^{inc}(\mathbf{r}) \\ \mathbf{E}_{PW}^{scat}(\mathbf{r}) 
\end{pmatrix}
 = E_0 \sum_{l} i^l \frac{2l+1}{l(l+1)} \left[\begin{pmatrix}  \mathbf{M}^{o \ (1)}_{l,1} \\ -b_l \mathbf{M}^{o \ (3)}_{l,1}\end{pmatrix}
 - i \ \begin{pmatrix} \mathbf{N}^{e \ (1)}_{l,1} \\ -a_{l} \mathbf{N}^{e \ (3)}_{l,1}\end{pmatrix} \right]  \label{eq:planewave_dual}
\end{equation}
where $E_0=\omega^2\mu_0 \frac{p_0}{4\pi r'} e^{ikr'}$, and $p_0$ is the amplitude of the dipole source given by $\mathbf{p(r')}=p_0 \delta(r-r')\hat{e}_x$, and for simplicity we normalise all our field results from Eq.~\ref{eq:planewave_dual} with $E_0$. 
Note that $r' \rightarrow \infty$ and the $x$-polarization of the plane wave enforces $m=1$, which reduces $C_{lm} s_{lm}$ and $-i C_{lm} t_{lm}$ to $i^l\frac{2l+1}{l(l+1)} \frac{p_0}{4\pi r'} e^{ikr'}$ (see Supp. Info. Section~3 for the full derivation). 
For a molecule at position $\mathbf{r}_0$ with a dipole moment along the unit vector $\hat{\mathbf{\mu}}$, the $\gamma_{exc}$ is given by:
\begin{equation}
\gamma_{exc} = \frac{\left| \hat{\mathbf{\mu}} \cdot \left[\mathbf{E}_{PW}^{inc}(\mathbf{r}_0) +\mathbf{E}_{PW}^{scat}(\mathbf{r}_0)\right] \right|^2}{\left| \hat{\mathbf{\mu}} \cdot \mathbf{E}_{PW}^{inc}(\mathbf{r}_0) \right|^2}  \label{eq:excitation_rate2}
\end{equation}
and is plotted in Figure~\ref{fig:IsoYYComp} (blue full lines) for an $x$-polarized molecule placed 0.5nm away from the NP surface along the $x$-axis---together with numerical FDTD calculations (blue dashed lines)---for two NP sizes of diameters 60nm and $1\mu$m.

The radiative decay rate ($\gamma_{rad}$) for the same spherical NP is obtained by placing a dipole source at position $\mathbf{r'}$ with a dipole moment $\mathbf{p}(\mathbf{r'})$ along the $x$-axis, and integrating the emitted energy on the surface of a putative sphere with radius $R=1m$ enclosing the overall system (NP and dipole source). 
It is then normalized to the energy emitted by the dipole source without the plasmonic environment present (see Supp. Info. Section~4 for further details), to produce the radiative decay rate of the plasmonic environment given by~\cite{Frezza2018}: 
\begin{equation}
\gamma_{rad}= \frac{\sum_L \left(2-\delta_0\right) C_{l,m} \left[ \left| s_{l,m}^{e,o \ (1)}(k\mathbf{r'}) -b_l \ s_{l,m}^{e,o \ (3)}(k\mathbf{r'})  \right|^2 + \left| t_{l,m}^{e,o \ (1)}(k\mathbf{r'}) -a_l \ t_{l,m}^{e,o \ (3)}(k\mathbf{r'})  \right|^2  \right] }{\sum_L \left(2-\delta_0\right) C_{l,m} \left[ \left| s_{l,m}^{e,o \ (1)}(k\mathbf{r'}) \right|^2 + \left| t_{l,m}^{e,o \ (1)}(k\mathbf{r'})  \right|^2 \right]}  \label{eq:radiative_decay_rate}
\end{equation}
where $\{ s_{l,m}^{e,o \ (1)},t_{l,m}^{e,o \ (1)} \}$ and $\{ s_{l,m}^{e,o \ (3)},t_{l,m}^{e,o \ (3)} \}$ describe how efficiently a dipole source at $\mathbf{r}'$ couples directly into free space and plasmonic modes of the system respectively, and $\{ a_l,b_l \}$ are the scattering Mie coefficients. 
When both $a_l$ and $b_l$ are zero, the dipole source no longer couples energy into the plasmonic modes, and the $\gamma_{rad}$ returns to unity. 
Eq.~\ref{eq:radiative_decay_rate} is plotted in Figure~\ref{fig:IsoYYComp} (red full lines) together with numerical FDTD calculations (red dashed lines) for comparison. 
For the data shown in Figure~\ref{fig:IsoYYComp}, the dipole source is placed 0.5nm away from the surface of the NP along the $x$-axis ($r_p$+0.5nm,0,0), and $\mathbf{p}(\mathbf{r}')$ is $x$-polarized. 
One can see that the analytical predictions and numerical calculations are in strong agreement for both NP sizes, showing that Eq.~\ref{eq:excitation_rate} and Eq.~\ref{eq:radiative_decay_rate} describe the system fairly well.
It is worth noting that despite the quantitative agreement between the numerical and analytical results, small differences originate from numerical limitations: the very close proximity of the dipole source/detection point to the surface of the NP requires extreme sub-nanometer meshing, and to increasing this any further would be too computationally expensive. 
The $\gamma_{rad}$ is affected much more severely by the meshing limitations than the $\gamma_{exc}$ due to the closeness of the dipole source to the NP(s), as the majority of the fields are concentrated in the high-meshed region. 
Figure~S3 compares the $\gamma_{exc}$ for a 60nm diameter NP using three different numerical methods, highlighting the variance with mesh type and numerical precision. 

Figure~\ref{fig:IsoYYComp} shows the comparison between the $\gamma_{exc}$ and $\gamma_{rad}$ for NPs with diameters of $60$nm and $1\mu m$, obtained both analytically from Eq.~\ref{eq:excitation_rate2} and Eq.~\ref{eq:radiative_decay_rate} as well as numerically with FDTD calculations. 
For the spherical gold $60$ nm NP, both the numerical and analytical results show that $\gamma_{exc}=\gamma_{rad}$. 
The multipolar decomposition reveals that only the first order ($l=1$) mode contributes to both the $\gamma_{exc}$ and $\gamma_{rad}$. 
Therefore the energy couples into the system in an identical manner as it out-couples, and always via $l=1$. 

\begin{figure}
\includegraphics[width=\linewidth]{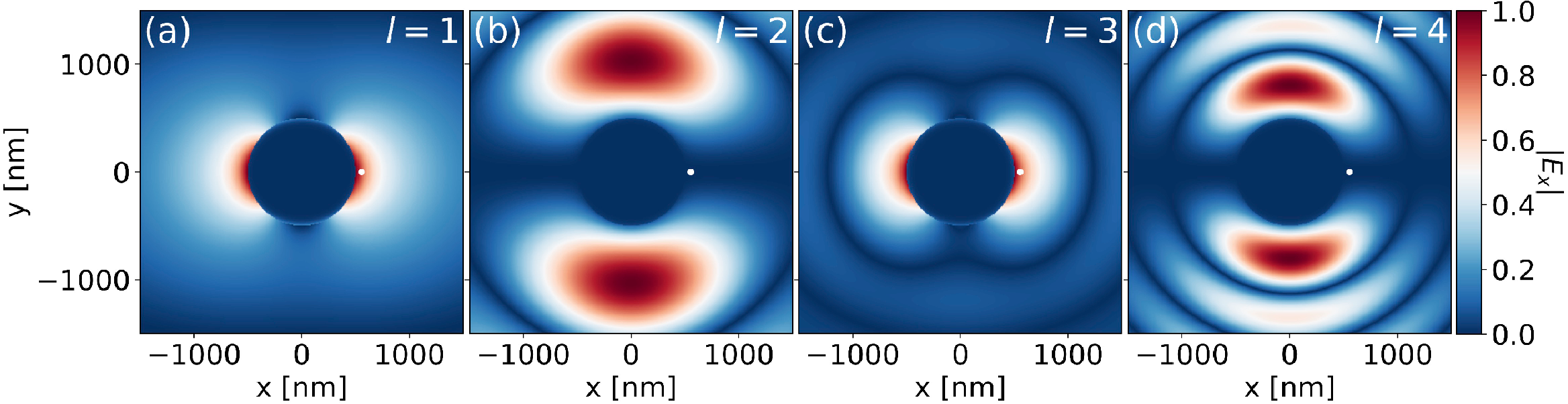}
\caption{\label{fig:PWSelectivity} Total $|E_x|$ fields of a $1\mu$m NP excited by an $x$-polarised plane wave, obtained from the multipolar decomposition for modes (a) $l=1$, (b) $l=2$, (c) $l=3$, (d) $l=4$. The white dot indicates the position where the $\gamma_{exc}$ is measured.}
\end{figure}

However, if one considers a larger spherical gold NP of diameter $2r_{p}=1$ $\mu m$, as shown in Figure~\ref{fig:IsoYYComp}(b), higher order modes become significant for both the $\gamma_{exc}$ and $\gamma_{rad}$---see Figure~S4 for the modal decomposition. 
This leads to significant differences between the $\gamma_{exc}$ and $\gamma_{rad}$.
This is because a dipole source placed close to the surface of the NP efficiently couples into every $l$-mode, each of which has multiple ($2l + 1$)-configurations for the plasmon fields, and all contribute to the $\gamma_{rad}$. 
A dipole source emits with multiple wavevectors along all directions and, as such, there is always a $(2l+1)$-configuration available for the dipole to couple maximally to (see Figure~S5). 
Therefore, a dipole source couples energy to each $(l,m)$-mode and radiates to the far-field with each $(l,m)$-mode according to its properties. 
However, the $\gamma_{exc}$ has fewer resonances, and the multipolar decomposition (see Figure~S4(a)) shows that only the odd $l$-modes are excited---which agrees with the numerical calculations. 
A plane wave incident on the structure has a certain polarisation and propagation direction (i.e. wavevector) that breaks the ($2l+1$) symmetry of the plasmonic modes, which fixes the value of $m$ (here the $x$-polarized plane wave defines $m=1$) and leads to $\pi/2$-rotations between consecutive $l$-modes~\cite{Babaze2020}. 
Therefore, this only allows for the maximal excitation of odd $l$-modes, as shown in Figure~\ref{fig:PWSelectivity}---which plots the $|E_x|$ fields for each $l$-mode of an isolated $1\mu$m NP due to an $x$-polarised plane wave excitation. 
The corresponding $|E_y|$ and $|E_z|$ components are shown in Figures~S6 and ~S7, respectively. 
The multipolar decomposition highlights that only the odd $l$-modes ($l=1,3,5,...$) contribute to the $\gamma_{exc}$ of a molecule placed at $\mathbf{r'}$, due to the field configuration of the modes excited by the plane wave.

\begin{figure}
\includegraphics[width=\linewidth]{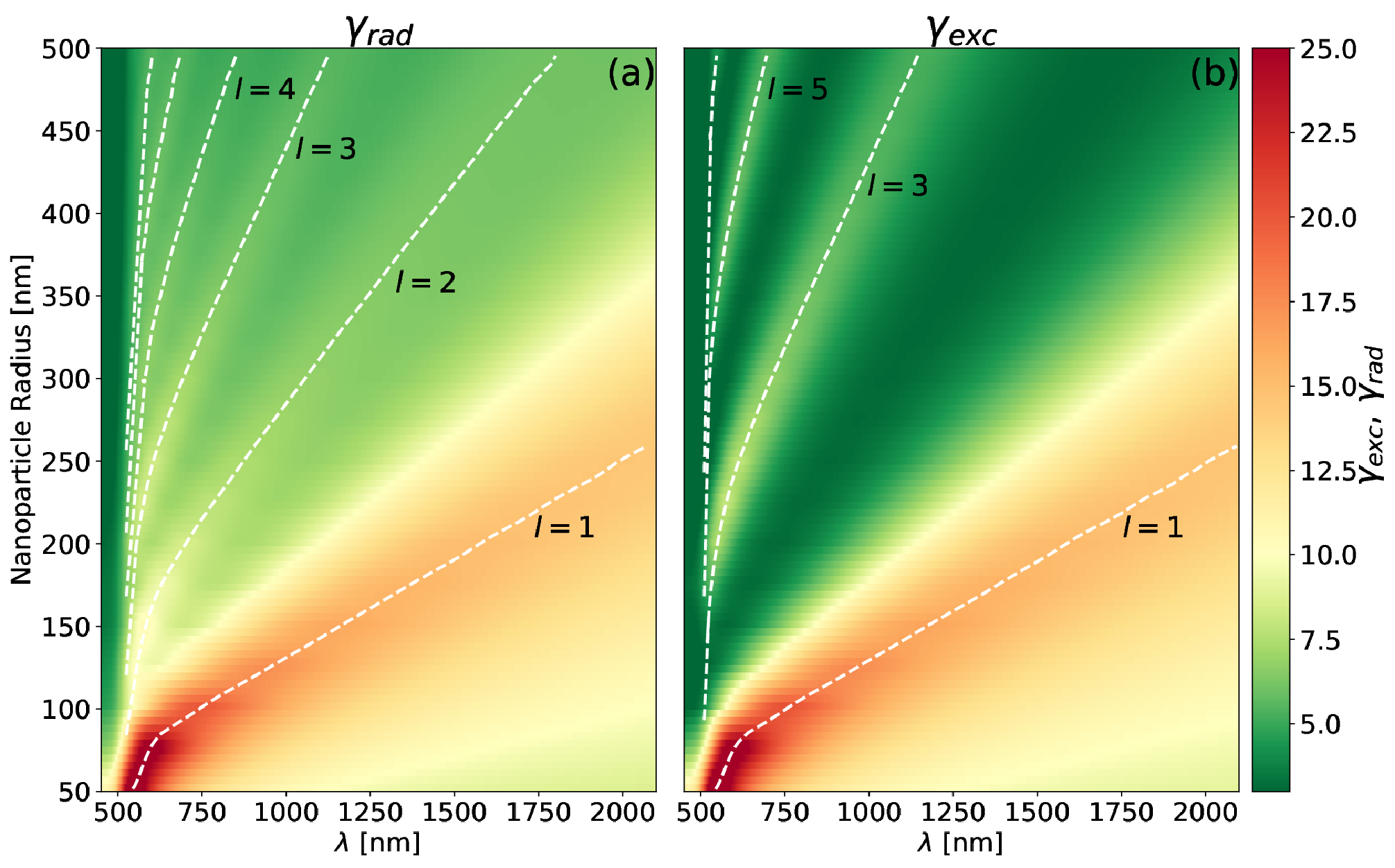}
\caption{\label{fig:YYRadDep}The (a) $\gamma_{rad}$ and (b) $\gamma_{exc}$ for isolated gold NPs obtained analytically from the multipolar decomposition, as a function of wavelength and the NP radius---ranging between $50-500nm$. The white dashed lines highlight the maxima curves for each $l$-labelled resonance.}
\end{figure}

To further understand the regime where the differences between the $\gamma_{exc}$ and $\gamma_{rad}$ become prominent, in Figure~\ref{fig:YYRadDep} we plot $\gamma_{exc}$ from Eq.~\ref{eq:excitation_rate2} and $\gamma_{rad}$ from Eq.~\ref{eq:radiative_decay_rate} for a full range of NP sizes. 
As expected, the larger the NP the greater the red-shift of the plasmonic modes~\cite{Benz2016}. 
However, it is immediately evident that only the odd $l$-modes contribute to the $\gamma_{exc}$, while the $\gamma_{rad}$ consists of all $l$-modes.
This behaviour becomes significant only for NP sizes beyond the quasi-static limit ($2r_{p}>150\,$nm), where the plane wave's phase propagation across the NP becomes significant. 
This is evident from Figures~S8 and~S9, where the $|E_x|$-fields of the first three $l$-modes are plotted for NP diameters within and beyond the quasi-static limit, for a plane wave and dipole source excitations, respectively. 
Note that electromagnetic reciprocity holds for all these systems, since the Green's dyadic has remained invariant ($\mathbf{\overleftrightarrow{G}} (\mathbf{r},\mathbf{r'})=\mathbf{\overleftrightarrow{G}} (\mathbf{r'},\mathbf{r})$) under the interchange of a source and a detector; Eq.~\ref{eq:Greens_dyadic} and Eq.~\ref{eq:scatter-field} are actually used to produce the analytic data of Figures~\ref{fig:IsoYYComp}-\ref{fig:YYRadDep}, which reveal the differences between the excitation and radiative decay rates. 
The differences arise from the fact that $\ s_{l,m}^{e,o \ (3)} (k\mathbf{r'})$ and $ -i t_{l,m}^{e,o \ (3)}(k\mathbf{r'})$ reduce to $\frac{p_0}{4\pi r'} i^l\frac{2l+1}{l(l+1)}$ for a plane wave. 
It is therefore evident that although nanoplasmonic systems obey electromagnetic reciprocity, it does not necessarily mean that energy in- and out-couples with equal rates. This leads to significant differences in how one excites QEs in plasmonic systems, and how to interpret experimental results measured in the far-field.
Hence, there are underlying differences in how energy is coupled into and out of plasmonic systems---especially beyond the quasi-static regime.

\section{\label{sec:CoupingComp}Plasmonic Nanoantennas}
\begin{figure}
\includegraphics[width=0.80\linewidth]{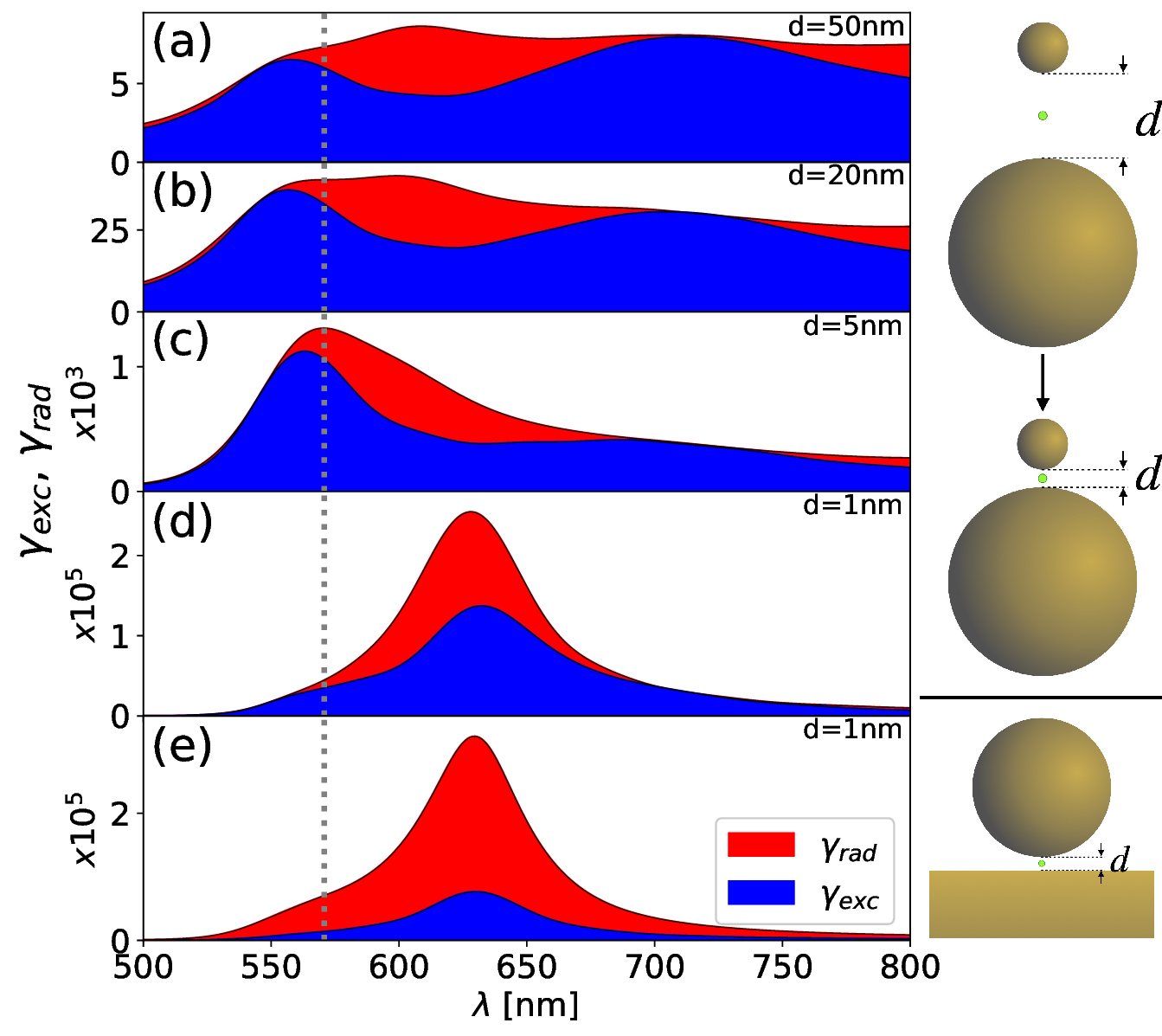}
\caption{\label{fig:YYSepDep}The $\gamma_{exc}$ (blue) and $\gamma_{rad}$ (red) calculated numerically for (a-d) asymmetric gold dimer antennas of diameters of 60nm and 1$\mu$m, and (e) the NPoM configuration with a gold NP of diameter $60$ nm. 
These have gap sizes of (a) $50$ nm, (b) $20$ nm, (c) $5$ nm, and (d-e) $1$ nm. 
The grey dashed line shows the $l=1$ resonance of the isolated $60$ nm gold NP. 
The right figure illustrates the closure of the gap, with the green dot indicating the position of the dipole source when calculating the $\gamma_{rad}$, and the location where the fields are measured when calculating the $\gamma_{exc}$. }
\end{figure}
Similar behaviour persists for plasmonic nanoantennas. 
By bringing two quasi-static plasmonic NPs close together, we form a dimer antenna that hybridises the modes of each NP.  
Within the quasi-static limit, the hybridisation simply comprises of the $l=1$ mode from each NP, and therefore quasi-static nanoantennas maintain equal $\gamma_{exc}$ and $\gamma_{rad}$---as shown in Figure~\ref{fig:NanoCavYComp}(b). 

If one couples together a quasi-static and a non---quasi-static NP, however, differences between the $\gamma_{exc}$ and $\gamma_{rad}$ emerge as shown in Figure~\ref{fig:YYSepDep}---which shows the $\gamma_{exc}$ (blue) and $\gamma_{rad}$ (red) for these nanoantennas, calculated using FDTD techniques.
Here we look at a plasmonic antenna formed by a small NP of diameter $60$ nm and a large NP of diameter $1\, \mu m$. 
Since we understand well how the $\gamma_{exc}$ and $\gamma_{rad}$ emerge for the two NPs separately, we start with a large separation where the response of the asymmetric dimer antenna is dominated by that of the $1\mu$m NP, and is nearly identical to Figure~\ref{fig:IsoYYComp}(b)--since the two NPs barely couple to each other. 
We then gradually couple them to form an antenna by reducing their separation from $50$ nm to $1$ nm. 
As the cavity size is reduced, the $l=1$ mode of the 60nm NP (that resonates at $550$nm, as shown with a grey dashed line) hybridises with the multiple modes of the $1\mu$m NP that exist within the same frequency regime. 
The coupling of the two NPs increases both the $\gamma_{exc}$ and $\gamma_{rad}$ by orders of magnitude. 
However, the differences between the $\gamma_{exc}$ and $\gamma_{rad}$ present for the $1\mu$m NP remain, and are in fact accentuated by the coupling of the two NPs. 
As the nanocavity approaches separations below $5$nm, the two NPs couple even tighter; the plasmonic resonances significantly red-shift, and even larger differences in the $\gamma_{exc}$ and $\gamma_{rad}$  emerge.  
Figure~\ref{fig:YYEqual}(a) and (c) show that the $l=1$ mode of the $60$ nm NP couples to the $l=5,6$ and $7$ modes of the $1$ $\mu m$ NP, since they spectrally overlap. 
The differences introduced to the $\gamma_{exc}$ and $\gamma_{rad}$ from the $l=6$ mode of the  $1$ $\mu m$ NP, lead to the differences observed in the combined system of the asymmetric dimer antenna. 
Therefore, the origin of the unequal coupling in nanoantenna systems can be directly traced back to contributions of the isolated components,
and more precisely to large NPs.

The NPoM antenna is qualitatively equivalent to the asymmetric dimer antenna shown in Figure~\ref{fig:YYSepDep}, but the large NP is now infinitely large. 
Although the mirror does not support discrete eigenmodes like the $1$ $\mu m$ NP does, it supports a continuum of evanescent eigenmodes (i.e. propagating surface plasmon polaritons)---these hybridize with the $l=1$ mode of the quasi-static NP in a similar manner to the asymmetric dimer antenna.
In Figure~\ref{fig:YYSepDep}(e) we plot the $\gamma_{exc}$ and $\gamma_{rad}$ for the NPoM with a NP diameter of $60$ nm and a gap size of $1$ nm which shows a very similar behaviour to the asymmetric dimer antenna with the same gap and small NP sizes. 
For the NPoM, the differences between the $\gamma_{exc}$ and $\gamma_{rad}$ are more prominent than the asymmetric dimer antenna, since the continuum of modes from the mirror leads to a greater number of even $l$-modes contributing to the $\gamma_{rad}$.
It is worth noting that the incident plane field here (parallel to the mirror) may not be experimentally feasible, due to the infinitely large size of the substrate. However, the response of any oblique angular excitation is a superposition of the normal and in-plane incidences, relative to the mirror, with the normal incidence not contributing to the same frequency regime---as shown in Figure~S10. 
The effect of the emitter position in this cavity is explored in Figure~S11-12, where we find that the relationship between the excitation and radiative decay rate is independent of the emitter's position.

\subsection{Tailoring the excitation and radiative properties of plasmonic nano-antennas}
\begin{figure}
\includegraphics[width=\linewidth]{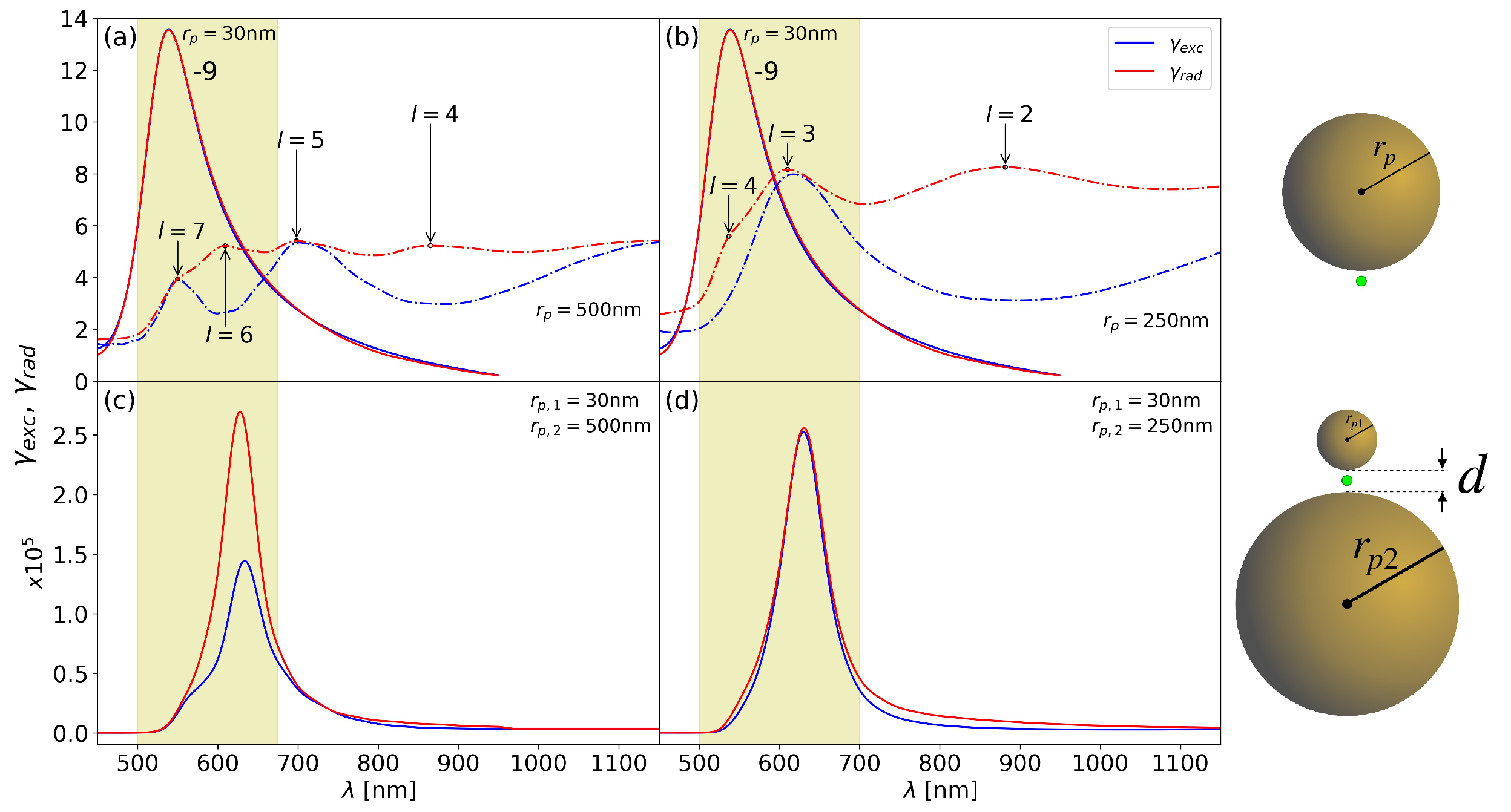}
\caption{\label{fig:YYEqual}The $\gamma_{exc}$ (blue) and $\gamma_{rad}$ (red) for: isolated gold NPs of diameters (a) $2r_p = 60$nm (full lines) and $2r_p = 1\mu$m (dashed lines), and (b) $2r_p = 60$nm (full lines) and $2r_p = 500$ nm (dashed lines); and for asymmetric gold dimer antennas with gaps of $d=1$ nm and diameters (c) $2r_{p,1} = 60$nm  and $2r_{p,2} = 1$ $\mu m$, and (d) $2r_{p,1} = 60$nm  and $2r_{p,2} = 500$ nm. The green dot in the right most figures indicates the position of the dipole source when calculating the $\gamma_{rad}$, and the location where the fields are measured for the $\gamma_{exc}$.}
\end{figure}
Since we now understand how the $\gamma_{exc}$ and $\gamma_{rad}$ emerge in plasmonic antennas, we aim to design a system with customised relative excitation and radiative properties. 
Although one can have more design flexibility by considering non-spherical plasmonic structures, here---to be consistent with the plasmonic systems we have shown so far---we only consider spherical dimer antennas. 
 
By consulting Figure~\ref{fig:YYRadDep} on the size dependence of NP's resonant frequencies and spectral separations, one can construct a dimer antenna with non---quasi-static elements that also offers $\gamma_{exc}=\gamma_{rad}$. 
We look for a spherical NP with an odd $l$-mode that spectrally overlaps with a well-separated odd $l$-mode of another sized NP; here we choose spherical NPs of diameters $60$ nm, with the $l=1$ mode resonant at 550nm, and 500nm, with the $l=3$ mode resonant at ~600nm. 
The $\gamma_{exc}$ and $\gamma_{rad}$ for these two NPs when they are isolated are plotted together in Figure~\ref{fig:YYEqual}(b). 
The $l=1$ mode of the $60$ nm NP spectrally overlaps with the $l=3$ mode of the $500$ nm NP. 
Although the $l=2$ and $l=4$ of the $500$ nm NP are spectrally close to the $l=3$ mode, the narrower bandwidth of the $l=1$ mode of the $60$ nm NP means that it primarily couples to the $l=3$ mode of the $500$ nm NP. 
The $\gamma_{exc}$ and $\gamma_{rad}$ for the coupled system is shown in Figure~\ref{fig:YYEqual}(d) for a 1nm gap, where one can see that $\gamma_{exc}\approx \gamma_{rad}$. 
Note that the small gap red-shifts the $l=1$ mode of the 60nm NP to spectrally overlap with the $l=3$ mode of the 500nm NP. 
Although the $500$ nm NP is well beyond the quasi-static limit, the fact that the $60$ nm NP couples primarily to a singular odd $l$-mode of the $500$ nm NP allows the system to retain $\gamma_{exc}\approx\gamma_{rad}$. 
Small differences arise in the frequency regime of the $l=2$ and $l=4$ modes of the isolated $500$ nm NP (i.e. $\lambda=700-1000$ nm and $\lambda=540-570$ nm respectively), due to their slight overlap with the tails of the 60nm NP's $l=1$ mode. 
This behaviour is in contrast to the coupling of the same $60$ nm NP with the modes of the larger $1$ $\mu m$ NP ($l=5,6,7$)---as we saw in Figure~\ref{fig:YYEqual} (a) and (c). 
Hence, to ensure equal in- and out-coupling of energy from a non---quasi-static plasmonic system one needs to primarily couple modes from each structure that are: of an odd-$l$ order, overlapping in frequency, and sufficiently narrow-band/spectrally separated from neighbouring even $l$-modes. 

The $\gamma_{exc}$ and $\gamma_{rad}$ of plasmonic nano-antennas respectively determine how efficiently one can excite an emitter within the antenna, and how efficiently a photon emitted by a molecule/quantum dot is transmitted to the far-field to be measured experimentally. 
Although it is often considered that the $\gamma_{exc}$ and $\gamma_{rad}$ are equal due to the reciprocal behaviour of electromagnetic systems, we have shown that this is not always the case when non---quasi-static structures are involved. 
This emerges from the polarization selection of the plasmonic modes excited on non---quasi-static structures by a plane wave. 
In general, plasmonic antennas with non-quasistatic elements out-couple energy much more efficiently than a plane wave can couple energy into the antenna and excite a QE ($\gamma_{rad}> \gamma_{exc}$)---even though we have shown that reciprocity holds (i.e. $\mathbf{\overleftrightarrow{G}} (\mathbf{r},\mathbf{r'})=\mathbf{\overleftrightarrow{G}} (\mathbf{r'},\mathbf{r})$). 
The relative difference between the $\gamma_{exc}$ and $\gamma_{rad}$ has very significant consequences for measurements where the near-to-far-field relationship is of high importance~\cite{Rose2014,Sugimoto2015,Baranov2018}. 
An example is that of surface-enhanced Raman spectroscopy (SERS), where a laser excites the chemical bonds of a molecule which then decays into one of the molecule's vibrational energy states, and leads to photon emissions measured in the far-field to produce Raman signals~\cite{FelixBenz2016,Long2002}. 
The intensity of the Raman signals changes with the properties of the plasmonic antenna. 
Similarly for the strong coupling of a few molecules at room temperature, where one excites the fluorescent molecules with a plane wave and observes the hybrid states via the radiative waves in the far-field~\cite{Chikkaraddy2016a,Wersall2017}. 
Possible future applications of such systems on the engineering of quantum states at room temperature with plasmonic nanoantennas, would have to account for the excitation and radiative properties of the plasmonic environment. 
Hence, it is vital to understand how the $\gamma_{exc}$ and $\gamma_{rad}$ emerge in plasmonic systems, and be able to design nanoplasmonic structures with the necessary properties.

\section{\label{sec:Conclusion}Conclusion}
In recent years, quantum plasmonics---where quantum emitters  (i.e. fluorescent molecules, quantum dots) are coupled to a plasmonic structure---has become a very promising photonic platform to bring quantum effects, observations and measurements at room temperature. 
The most prominent and commonly used plasmonic nanostructures are the dimer antenna and the NPoM configuration. 
However, until now it was believed that one can in- and out-couple energy equally in such systems, which we show is not valid for all plasmonic systems. 
We use a multipolar decomposition model (Mie theory) to decompose the modes of isolated NPs, and reveal the contribution of each mode to the excitation rate, $\gamma_{exc}$,  (i.e. in-coupling) and radiative decay rate, $\gamma_{rad}$, (i.e. out-coupling).
We find that for non---quasi-static plasmonic systems, the even $l$-modes only contribute to $\gamma_{rad}$. 
Therefore, the radiative energy from an emitter placed at close proximity to the plasmonic structure is larger than the excitation of the same emitter from a plane wave. 
This behaviour persists for coupled plasmonic systems, such as plasmonic antennas with non---quasi-static elements (i.e the NPoM). 
Finally, we show how to design antennas that have tailored relative $\gamma_{exc}$ and $\gamma_{rad}$. 
This study unveils how to create plasmonic antennas for applications where the near-to-far-field relationships is very important, such as: SERS~\cite{FelixBenz2016}, the strong coupling of a few molecules with plasmons at room temperature~\cite{Chikkaraddy2016a}, and other quantum plasmonic applications, such as quantum computing with DNA-origami controlled qubits~\cite{Chikkaraddy2018}.

\section{\label{sec:Methods}Methods}
The description of the relative electric permittivity for gold used throughout the analytical description is fitted  to the Johnson and Christy experimental data for gold \cite{Johnson1972}, and follows the Drude-Lorentz model:
\begin{eqnarray}
\varepsilon = \varepsilon_{\infty} - \frac{\omega_p^2}{\omega^2 + i\gamma\omega} + \frac{\sigma_1 \omega_{p,1}^2}{\omega_{p,1}^2 - \omega^2 - i\gamma_1 \omega} + \frac{\sigma_2 \omega_{p,2}^2}{\omega_{p,2}^2 - \omega^2 - i\gamma_2 \omega}
\end{eqnarray}
where $\omega$ is the angular frequency; $\varepsilon_\infty = 4.9752$ is a constant relative electric permittivity; $\sigma_1 = 1.76$ and $\sigma_2 = 0.952$ are the strengths of the two Lorentz oscillations; $\omega_p = 1.345\times 10^{16}$s$^{-1}$, $\omega_{p,1} = 1.774\pi \times 10^{15}$s$^{-1}$ and $\omega_{p,2} = 1.372\pi \times 10^{15}$s$^{-1}$ are respectively the plasma frequencies for the Drude term, and the first and second Lorentz oscillations; and similarly for the Drude and Lorentz oscillation damping coefficients $\gamma = 1.839\pi \times 10^{13}$s$^{-1}$, $\gamma_1 = 6.338\pi \times 10^{14}$s$^{-1}$ and $\gamma_2 = 3.564\pi \times 10^{14}$s$^{-1}$. 

FDTD calculations were performed using Lumerical FDTD Solutions software~\cite{FDTD}. The electric permittivity for gold is fitted to the  Johnson and Christy experimental data for gold \cite{Johnson1972}. 
Throughout these calculations, all nanocavity systems use 1nm separation, with the dipole source placed at the centre of the cavity. 
To keep this placement consistent in all non-cavity systems, the dipole source was placed at the same $0.5$nm distance from the surface of the NP. 
All plane waves considered here are $x$-polarised and propagating along the positive $z$-axis, and all dipole sources are $x$-polarised. 
Very fine meshing of $0.9$ nm is applied to the $60$ nm NPs, and a harsher meshing of up to $10$ nm for the largest systems ($2r_p = 1$ $\mu m$) due to the increased computational demand. 
Finer meshing of $0.1$ nm is applied within the nanocavity region, where the greatest field enhancements are produced. In every FDTD simulation, 12 layers of PMLs were used to minimise the effects of PML reflections on our results. 
In addition, the simulation domain was kept proportionally constant at $20 r_p$---the radius of the largest NP in the system---to insure a proper convergence of the simulations. 

Additionally, COMSOL MultiphysicS has been used to compare the analytical results of Mie theory shown in Figure~\ref{fig:NanoCavYComp} with numerical calculations. 
We chose to run these calculations with COMSOL as we can define the electric permittivity of gold with the same Drude-Lorentz model used analytically (Lumerical FDTD does not allow for analytical description of gold); it also enabled us to check the FDTD inaccuracies emerging from meshing a spherical NP with Yee cells. 
This ensured that differences between the analytical and numerical COMSOL results were due to numerical errors alone.

\section*{Acknowledgements}
AD gratefully acknowledges support from the Royal Society University Research Fellowship URF\textbackslash R1\textbackslash 180097, Royal Society Research Fellows Enhancement Award RGF \textbackslash EA\textbackslash 181038, Royal Society Research grants RGS \textbackslash R1\textbackslash 211093 and funding from EPSRC for the CDT in Topological Design EP/S02297X/1.

\end{document}